\documentstyle[preprint,eqsecnum,aps]{revtex}
\begin{document}
\draft
\title{Buckling Instabilities of a Confined Colloid
Crystal Layer}
\author{T. Chou and David R. Nelson}
\address{Dept. of Physics, Harvard University, Cambridge, MA 02138}
\date{\today}
\maketitle
\begin{abstract}
\baselineskip7mm A model predicting the
structure of repulsive, spherically symmetric,
monodisperse particles confined between two
walls is presented. When plate separations are
small, only one layer of particles can be
confined; however, when the plate separation is
increased, multiple layers will eventually
form. We study the buckling transition of a
single flat layer as the double layer state
develops. Experimental realizations of this
model are suspensions of stabilized colloidal
particles squeezed between glass plates. By
expanding the thermodynamic potential about a
flat state of \( N \) confined colloidal
particles, we derive a free energy as a
functional of in-plane and out-of-plane
displacements. As the gap separation increases,
certain out-of-plane modes soften. The
wavevectors of these first buckling
instabilities correspond to three different
ordered structures. Landau theory predicts that
the symmetry of these phases allows for second
order phase transitions. This possibility
exists even in the presence of gravity or plate
asymmetry. These transitions lead to critical
behavior and phases with the symmetry of the
three-state and four-state Potts models, the
X-Y model with 6-fold anisotropy, and the
Heisenberg model with cubic interactions.
Experimental detection of these
structures is discussed.
\end{abstract}
\vspace{5mm}

\section{Introduction}
Colloid suspensions have long been used in
practical applications such as paints, coatings
and many manufacturing processes \cite{FCS}.
Recently, new uses of colloids such as in
electro-optic devices and in semiconductors have
been discovered \cite{CEO}.  These new uses, as
well as the general understanding of colloidal
ordering applicable in other fields, require a
thorough knowledge of the structures and phases
present in these suspensions.  Hence, colloid
structure and rheology have been under intense
study across many disciplines \cite{CEO}.

Colloid particles, usually on the order of $1\mu
m$ in diameter, are a convenient size for
laboratory study. Due to the complicated
interactions these particles experience, colloid
suspensions exhibit a rich cornucopia of phases
in solution. These interactions can be described
by electrostatic, van der Waals, hydrodynamic
and steric effects.  When repelling moieties
such as polymers or charged molecules are
attached to the surfaces, the colloids are
stabilized against aggregation and extended
ordered structures in solution can form.
Debye-Huckel theory is typically used to model
the screened electrostatic interactions in a
charge stabilized suspension. Electrostatic
effects, together with van der Waals and hard
sphere interactions, are important ingredients
in the Derjaguin-Landau-Verwey-Overbeek (DVLO)
theories\cite{FCS}.

In bulk, monodisperse colloidal systems can
appear in liquid, crystalline, amorphous, and
inhomogeneous phases. Crystals with  bcc, fcc,
and random stacked close packed structures have
been observed.

Theoretically, the ordering of hard or otherwise
repulsive spheres have been approached
analytically and via computer simulations. Much
effort has been directed towards understanding
the nature and consequences of DVLO pair
potentials. A Yukawa interaction was found to
give rise to structural features consistent with
observations\cite{KRG}.  Pair potentials have
also been used to described the dynamics of
colloid crystals\cite{JPN}.

Experimentally, these systems have been studied
with microscopy, video imaging, and light
diffraction\cite{CAM2,PP2,PP4}. In bulk,
monodisperse colloidal systems can appear as
liquid, crystalline, amorphous, and
inhomogeneous states. Crystals with  bcc, fcc,
and random stacked close packed structures have
been observed\cite{FCS,TO}.

It was quickly realized that colloid suspensions
are an ideal system to use for studying two
dimensional phase transitions. The first
observation of 2D ordering was the interfacial
colloidal crystal of Pa.  Pieranski\cite{PP1}.
In this system, colloid particles floating on an
air-water interface interact mainly through
dipole-dipole forces. Subsequent studies have
involved immersed colloids with regions where
two plates are brought close together. The
plates are immersed in bulk solution allowing
for free expansion of the confined structures.
In addition, by using materials such as
poly(styrene), which is almost density
matched to the aqueous solvent, the effects of
gravity are negligible. In these studies, the
bulk solution provides a fixed chemical
potential and thermodynamic particle resevoir
for the confined sample.

One class of experiments for which confined
colloidal systems are particularly suited is the
study of melting in two dimensions\cite{CAM2}.
Because of their macroscopic size, colloid
spheres do not feel the atomic scale granularity
of the confinement walls, (see Fig 1). Thus,
this system is an ideal candidate for observing
dislocation and disclination driven melting
mechanisms in 2D.  In fact, the search for
defect mediated melting appears to be most
successful in these confined colloidal systems.
An isotropic fluid exists at narrow wall
separations; as $h$ is increased, the one layer
areal density increases. In practice, $h$ is
difficult to accurately control and a wedge
geometry with gradually increasing plate
separation is used. The correlations of particle
positions along this wedge are consistent with a
two stage dislocation/disclination mediated
melting process.

If $h$ is increased further, the continual
invasion of particles into the gap is allowed
only by formation of multiple layers. The
observations of many experimenters\cite{CAM2,PP2}
can be summarized by,

\begin{equation}
0\rightarrow fluid \rightarrow 1\triangle
\rightarrow 2\Box \rightarrow 2\triangle
\rightarrow \cdot\cdot\cdot N\Box \rightarrow
N\triangle \rightarrow (N+1)\Box \label{seq}
\end{equation}

\noindent where the gap $h$ is increases from
left to right and $\triangle $ and $\Box$
represent triangular and square lattices
within each of the $N$ layers. In the gap,
packing constraints dictate whether the layers
are square or triangularly ordered\cite{PP3}.
The $\Box$ and $\triangle$ layers are
approximately the (001) and (111) planes of
bcc and fcc lattices respectively. As these
transitions proceed, Murray et. al.
\cite{CAM2} have observed that large
fluctuations of the colloids perpendicular to
the plane occur; such fluctuations are
especially prevalent in the $1\triangle
\rightarrow 2\Box$ transition. As $N$
increases, the structure of the confined
layers approaches that of a bulk solution. In
this study, we use static colloidal particle
interactions to model the free energy of the
assembly as a function of particle
displacements. Stability conditions are
imposed and minimization based on the Landau
theory of structural phase
transitions\cite{SPT,LT} is used to predict
the possible buckled structures in the $ 1
\triangle \rightarrow 2\Box$ region.  The free
energy depends upon both the amplitude and
phase of these ``buckling waves''.

In the next section, we derive from
microscopic interactions the free energy of
the confined colloid system. Stability of this
model and the critical wave vectors are
examined in Section III. The phase
transitions into each of the possible states
is related to continuum statistical models in
Section IV. Structure factors and other
experimental consequences of these structures
are then briefly discussed in Section V.
Finally, a model for particles trapped
in a tubular pore is presented in the Appendix.

\section{Model Free Energy}
In this section, we derive a free energy
functional from the thermodynamic potential of
a confined layer of particles immersed in a
bulk particle resevoir.  The functional is
written in terms of particle displacements and
cast into the Landau-Ginzburg form. The
parameters in this expansion are defined by
the microscopic interactions in the problem.

For a confined collection of $N$ particles
mutually interacting via a spherically
symmetric potential $U(r)$, the free energy
can be written as a sum over particles $i$ and
$j$,

\begin{equation}
\Omega= \sum_{i<j} U (\vert \vec{r}_{i}
-\vec{r}_{j}\vert ) +
V(f_{i},h) -\mu N
\end{equation}

\noindent where

\begin{equation}
V(f,h) = V_{1}({h \over 2}-f)+
V_{2}({h \over 2}+f) + V_{ext}(f) \label{V0}
\end{equation}

\noindent is the total particle-plate
potential. $V_{1}$ and $V_{2}$ are the
individual plate-wall potentials, which we
usually assume equivalent when the
confining plates are identical. The term
$V_{ext}(f)$ describes the effects of
external fields such as gravity.  In that
case $V_{ext} \propto f$.  The ${\bf \hat{z}}$
component of $\vec{r}_{i}$ is denoted by $f_{i}$,
the height of particle $i$ measured from the
minimum of $V_{1}+V_{2}$.

For a single layer of spheres squeezed between
the plates just prior to buckling, the mean
particle positions form a flat triangular
lattice with lattice constant $a$. Although in
two dimensions there is only quasi-long range
translational order\cite{DRN}, we consider a
finite sample that is locally crystalline.

Assuming our $N$ particle sample is within a
well ordered domain, the sum over all
nearest neighbor interactions becomes,

\begin{equation} {\Omega_{o} \over A_{o}} =
{3NU(a_{1})+3NU(a_{2}) + \ldots -\mu
N+V(0,h)N \over N(a^2\sqrt 3/2)} \label{E1}
\end{equation}

\noindent where $A_{o} = Na^2 \sqrt {3}/2$ is
the area occupied by the undistorted $N$
particle lattice. The last term varies with
position in the wedge geometry of \cite{CAM2}:
this experimental arrangement is equivalent to
slow spatial variation in the chemical
potential. The second and further terms on the
right result from summing interparticle
energies from next and further neighbor
particles respectively; hence, $a_{1}\equiv a,
a_{2} = a \sqrt 3 ,$ etc.  For (\ref{E1}) to
be useful, the further neighbor interactions
must be truncated at a suitable distance much
smaller than the $N$ particle system size.
Upon minimizing $\Omega_{o}/{A}_{o}$
with respect to $a$, we find the equilibrium
lattice spacing and the chemical potential
associated with it.

The lattice spacing $a$ is uniquely determined
by the chemical potential provided $U(r)$ is
smooth and has at most one minimum. However,
further neighbor interactions and interactions
with multiple minima could lead to phase
transitions among flat structures of varying
lattice constants $a$.  Multiple minima are
often predicted by the various contributions
to interparticle potentials. We will assume
that the repulsive part of the potential is
strong enough to prevent aggregation and that
there are no in-plane lattice distortions near
the buckling transition when $f$ becomes
nonzero. The necessary requirements on $U(r)$
will be determined below.

Upon expanding the total free energy about the
flat phase $f=0$ and about an equilibrium
spacing $a=\vert {\bf x}_{i} - {\bf x}_{j}
\vert $ between nearest neighbors $i,j$, with
the definition of the in-plane displacement
field \( {\bf u}({\bf x}_{j+i}) - {\bf u}({\bf
x}_{j}) + {\bf a}_{i}^{n} = {\bf x}_{i} - {\bf
x}_{j} \), we have

\begin{equation}
\Omega = \sum_{\bf x} \sum_{n} \sum_{i=1}^{3}
U(\sqrt {\vert{\bf u}({\bf x}) - {\bf u}({\bf x}
+{\bf a}_{i}^{n}) - {\bf a}_{i}^{n} \vert^2 +\vert
f({\bf x})-f({\bf x}+{\bf a}_{i}^{n}) \vert
^2}\,\,) + \sum_{x} V(f({\bf x})) - \mu(a) N
\label{E2}
\end{equation}

\noindent Here, $\sum_{\bf x}$ is a discrete
sum over the sites of a triangular lattice and
$n$ indexes different coordination shells. The
$\{ {\bf a}_{i}^{n} \}$ are the three vectors
$\{i\}$ spanning the triangular lattice and
inclined at 120$^o$ to each other, each with
length $a_{n}$, as depicted in Fig. 2.  For
nearest neighbors,

\begin{equation}
{1 \over a_{1}}\{{\bf a}_{i}^{1}\} \equiv \{{\bf
e}_{i}^{1}\} \equiv \{ {\bf \hat{y}}, \, {\sqrt{3}
\over 2} {\bf \hat{x}}-{1 \over 2}{\bf \hat{y}}, \,
-{\sqrt{3} \over 2} {\bf \hat{x}}-{1 \over 2}{\bf
\hat{y}} \}.
\end{equation}

In order to write the free energy as a sum over a
fixed number of particles, we consider a
homogeneous system and a bulk chemical potential
which, together with $V$, determines the lattice
spacing and fixes the density of particles.

\subsection{Symmetric particle-wall potentials}
Upon expanding $\bar{\Omega} \equiv \Omega /A$
about the height $f$=0 and the in-plane
displacement ${\bf u} =0$, and using $\mu(a)$
calculated to the appropriate order of further
neighbor interactions, the nearest neighbor free
energy density becomes,

\begin{eqnarray}
\bar{\Omega} = const. +{K_{2} \over A}\sum_{\bf
x}\sum_{i=1}^{3} \vert {\bf e}_{i}
\cdot\Delta_{i}{\bf u} ({\bf x})\vert ^2 + {K_{1}
\over 2A} \sum_{\bf x}\sum_{i=1}^{3} \left( \vert
\Delta_{i} {\bf u}({\bf x}) \vert ^2 + \vert
\Delta_{i} f({\bf x})\vert ^2 \right) - \nonumber \\
{K_{2} \over 2aA}\sum_{\bf x}
\sum_{i=1}^{3}({\bf e}_{i} \cdot\Delta_{i} {\bf
u}({\bf x})) \vert\Delta_{i} f({\bf x})\vert ^2
+{K_{2} \over 8a^{2}A}\sum_{\bf x}
\sum_{i=1}^{3} \vert\Delta_{i} f({\bf x})
\vert ^4 + \nonumber \\
{K_{3} \over 48a^{4}A} \sum_{\bf x}\sum_{i=1}^{3}
\vert \Delta_{i}f({\bf x})\vert ^6 + \dots +
V(f,h) \label{E3}
\end{eqnarray}

\noindent with
\begin{equation}
V(f,h) =\sum_{\bf x}\left( {r\over 2} f^2({\bf x})
+ u f^4({\bf x}) + v f^6({\bf x}) + \ldots \right)
\end{equation}

\noindent and

\begin{equation}
A = {A_{o}\over 3N} \sum_{\bf x} \sum_{i=1}^{3}
 \left[1-{2\over a}({\bf e}_{i}^{1}\cdot
\Delta_{i}^{1}{\bf u}) +
{1 \over a^{2}} \vert \Delta_{i}^{1} {\bf u} \vert ^{2}
-{8 \over 3a^{2}} ({\bf e}_{i}^{1}\cdot
\Delta_{i}^{1}{\bf u})^{2} +{8 \over 3a^{2}}
({\bf e}_{i}^{1}\cdot \Delta_{i}^{1}{\bf u})
({\bf e}_{i+1}^{1}\cdot \Delta_{i+1}^{1}{\bf u})
 \right], \label{AREA}
\end{equation}

\noindent the projected area of $N$ colloid
particles to second order in ${\bf u}$. In these and
subsequent equations, objects with direction indices
$\{i\}$ are cyclic: $O_{i} \equiv O_{i+3}$.  The
notation \( {\bf a}_{i}^{1}\equiv {\bf a}_{i} \) and
\( K^{1}_{1,2,3} \equiv K_{1,2,3} \) corresponding
to a nearest neighbor approximation is used. In
going from (\ref{E2}) to (\ref{E3}), terms linear in
${\bf u}$ have been cancelled by $ \mu (a)/A$.  For
further neighbors, we use the notation \(
\Delta_{i}^{n}\phi \equiv \phi({\bf x}) - \phi({\bf
x}+ {\bf a}_{i}^{n}) \). To include the effects of
the $n^{th}$ nearest neighbor spheres at in-plane
distance $a_{n}$, additional
terms such as ${K_{1}^{(n)} \over 2A} \sum_{i,{\bf
x}} \vert \Delta_{i}^{(n)}f({\bf x}) \vert ^{2} $
must be added to Eq. (\ref{E3}). The constant term will
be omitted in future equations.

The parameters $K_{1}^{n}, K_{2}^{n}$, and $K_{3}^{n}$
are completely defined by the microscopic
particle-particle interactions,

\begin{equation}
K_{1}^{(n)} \equiv {1 \over a_{n}}{\partial U \over
\partial \xi}(a_{n}) \label{K1}
\end{equation}

\begin{equation}
K_{2}^{(n)} \equiv {\partial^{2}U \over \partial
\xi^{2}} (a_{n}) - K_{1}^{(n)} \label{K2}
\end{equation}

\begin{equation}
K_{3}^{(n)} \equiv {a_{n} \over 3}
{\partial ^3 U \over \partial
\xi^{3}} (a_{n}) - K_{2}^{(n)} \label{K3}
\end{equation}

The expression for $V$ is a Taylor expansion of
the total particle-plate potential.  In the
absence of an external field and using identical
confining plates, the function $V(f)$ is
symmetric and its expansion contains only even
powers of $f$.  The terms in equation (\ref{E3})
contain the contributions from competing
interactions. For repulsive particle-particle
interactions $K_{1}^{n} < 0$; consequently the
energy density maximizes \(
\vert\Delta_{i}f({\bf x}) \vert ^{2} \)
throughout sample. However, when repulsion from
the walls, $V(h,f)$, is strong enough, buckling
is prevented.

\subsection{Asymmetric Potentials}

Thus far, we have considered only symmetric
particle wall potentials in the absence of
external perturbations. In general however,
an external field may couple linearly
to the height function $f$,

\begin{equation}
V_{ext}(f) = \lambda \sum_{x}f({\bf x}) \label{V1}
\end{equation}

\noindent These perturbations may result
from the effects of gravity or plate material
asymmetry. An externally applied electric field
would couple in this manner but would also
induce dipoles resulting in an additional
repulsive force between the particles.
The total wall-particle potential becomes,

\begin{equation}
V(f,h) = {r(h) \over 2}f^2 + uf^{4} + vf^{6} +
\ldots + \lambda f \label{V2}
\end{equation}

In mean field theory, this linear term shifts the
minimum of each particle from $f=0$ to $\bar{f}$
where $\bar{f}$ is found by minimizing (\ref{V2})
with respect to $f$. Keeping terms only to fourth
order in $f$, we find,

\begin{equation}
\bar{f} \simeq {(3u)^{1/3}\left[ \sqrt
{9\lambda^2+r^3/3u} - 3\lambda \right]^{2/3}-r
\over 2(3u)^{2/3} \left[ \sqrt {9\lambda^2+r^3/3u}
-3\lambda \right] ^{1/3}}. \label{shift1}
\end{equation}

\noindent Re-expansion of (\ref{V0}) about
$\bar{f}$ yields\cite{AA},

\begin{equation}
V(f,h) \simeq const. + \sum_{\bf x} \left(
{\bar{r} \over 2}\eta({\bf x}) ^2
+ {\hat{\lambda}} \eta^{3}({\bf x}) +
u \eta^{4}({\bf x}) +\ldots \right). \label{V3}
\end{equation}

\noindent Here, $\eta = f-\bar{f}$, and the new
expansion parameters are related to the ones in
symmetric potential case by,

\begin{equation}
\bar{r} = r+12u\bar{f}^2
\end{equation}

\begin{equation}
\bar{\lambda} = 4u\bar{f}.
\end{equation}

\noindent When asymmetry is taken into account,
(\ref{E3}) is modified by replacing $f$ with
$\eta$, $r$ by $\bar{r}$ and adding the term
$\bar{\lambda} \eta^{3}$.  This cubic term
changes the nature of the transitions.

\section{Stability Analysis}
We now examine the region of validity of the
free energy (\ref{E3}). In particular, we
determine when and if phonons in the ${\bf u}$
or $f$ fields become  unstable or ``soft''. This
analysis is most convenient in Fourier space. We
define the Fourier modes,

\begin{equation}
{\bf u}({\bf x}) = \sum_{\bf k} {\bf u}
({\bf k})exp(ik{\bf x})
\end{equation}

\noindent and

\begin{equation}
f({\bf x}) = \sum_{\bf k} f({\bf k})exp(ik{\bf x})
\end{equation}

\noindent where $f({\bf k}) = f^{*}({\bf -k})$
and ${\bf u(k)}= {\bf u}^{*}({\bf -k})$ since the
displacements are real. When periodic boundary
conditions apply, the sums are over wavevectors
spaced $2 \pi /L$ apart where L is the linear
dimension of of the $N$ particle system.

To quadratic order, the free energy becomes

\begin{equation}
\bar{\Omega} \equiv {1\over 2NA}
\sum_{\bf G,k} \left[r(h)+\sum_{n} D_{n}({\bf k})
\right] f({\bf k}) f({\bf G-k}) + {1\over
2NA}\sum_{\bf G,k}{\bf
u(k)}\hspace{-1mm}\cdot\hspace{-1mm} \left(
\sum_{n} {\bf M}_{n}({\bf
k})\right)\hspace{-1mm}\cdot\hspace{-1mm} {\bf
u}({\bf G}-{\bf k}) \label{M0}
\end{equation}

\noindent In (\ref{M0}), we keep only the pieces
quadratic in ${\bf u}$ which are not surface
terms as they do not affect the eigenvalues
$\omega^{2}({\bf k})$ of this dynamical
matrix. The {${\bf G}$} are reciprocal lattice
vectors with $\vert {\bf G} \vert = 4\pi/
\sqrt{3}a$.

In this paper, we consider buckling
instabilities in out-of-plane modes with the
modes in ${\bf u}$ remaining stable. For the
in-plane phonons, stability requires that the
eigenvalues of the dynamical matrix, \(\sum_{n}
{\bf M}_{n}({\bf k}) \), be real and positive.
In the nearest neighbor approximation,

\begin{equation}
{\bf M}_{1}({\bf k}) = \left( \begin{array}{cc}
D_{1}({\bf k}) + {3 \over 2}K_{2} (2-cos\,{\bf
k}\hspace{-1mm}\cdot\hspace{-1mm}{\bf a}_{2}
-cos\,{\bf k}\hspace{-1mm}\cdot\hspace{-1mm}{\bf
a}_{3}) & {\sqrt 3 \over 2}K_{2} (cos\,{\bf
k}\hspace{-1mm}\cdot\hspace{-1mm}{\bf a}_{2}
-cos\,{\bf k}\hspace{-1mm}\cdot\hspace{-1mm}{\bf
a}_{3}) \\ {\sqrt 3 \over 2}K_{2} (cos {\bf
k}\hspace{-1mm}\cdot\hspace{-1mm}{\bf a}_{2}
-cos\,{\bf k}\hspace{-1mm}\cdot\hspace{-1mm}{\bf
a}_{3}) & D_{1}({\bf k}) + {K_{2} \over
2}(6-4cos\,{\bf k}\hspace{-1mm}\cdot\hspace{-1mm}{\bf
a}_{1} -cos\,{\bf k}\hspace{-1mm}\cdot
\hspace{-1mm}{\bf a}_{2}-cos\,{\bf k}\hspace{-1mm}
\cdot\hspace{-1mm}{\bf a}_{3})
\end{array} \right) \label{M1}
\end{equation}

\noindent where

\begin{equation}
D_{1}({\bf k}) = 2K_{1}(3-cos \, {\bf k\hspace{-1mm}
\cdot\hspace{-1mm} a}_{1} -cos \, {\bf
k\hspace{-1mm} \cdot\hspace{-1mm} a}_{2} - cos
\,{\bf k \hspace{-1mm}\cdot\hspace{-1mm} a}_{3})
\label{D1}
\end{equation}

We find that the all eigenvalues of ${\bf
M}_{1}({\bf k})$ are always real and are
positive provided,

\begin{equation}
K_{1}+{K_{2} \over 2} > 0 \label{C1}
\end{equation}

and

\begin{equation}
\left( {1 \over K_{2}} + {1 \over 2K_{1}} \right) ^2
> {1 \over D_{1}^{2}({\bf k})} \sum_{i=1}^{3} \left(
cos^2 \, {\bf k\hspace{-1mm}\cdot\hspace{-1mm} a}_{i}
- cos \, {\bf k\hspace{-1mm}\cdot\hspace{-1mm}
a}_{i}\,\,cos \, {\bf
k\hspace{-1mm}\cdot\hspace{-1mm} a}_{i+1} \right)
\label{C2}
\end{equation}

\noindent with ${\bf a}_{i}\equiv {\bf
a}_{i+3}$.  The criterion (\ref{C1}) is
satisfied with repulsive potentials in general.
In the first Brillouin zone, the right hand side
of (\ref{C2}) varies between 0 and
$K_{1}^{-2}/16$. Therefore in the nearest
neighbor approximation, in-plane phonons are
stable provided,

\begin{equation}
\left( K_{1} + {K_{2} \over 4} \right)
\left( K_{1} + {3K_{2} \over 4} \right) > 0
\end{equation}

\noindent This is just a more stringent version of
(\ref{C1}) and satisfied by potentials which are
repulsive and short ranged consistent with the
nearest neighbor approximation.

Now, we find instabilities in the magnitude of
$f({\bf k})$ and assume that the phase of the
displacement functions have been fixed to
minimize the free energy. The dependence of the
corrugations on the phase of $f$ will be
discussed later. Provided (\ref{C2}) holds, we
can consider only the instabilities in the
transverse modes, $f$.  Here, we require

\begin{equation}
\omega^{2}({\bf k}) \equiv r(h)+\sum_{n}D_{n}({\bf k}) \leq 0
\label{w}
\end{equation}

\noindent for {\it instability} of the ${\bf
k}^{th}$ mode of $f({\bf k})$. As the spacing
between the plates increases, the compressional
forces balancing the inward forces from the
colloid bath and $r(h)$ decrease. The points at
which the criterion (\ref{w}) is first satisfied
selects the first modes to soften. An estimate
of the critical spacing $h_{c}$ can be made by
solving \(r(h_{c}) \simeq -\sum_{n}D_{n}({\bf
k}_{c})\).

In the nearest neighbor truncation, \( r(h) =
-D_{1}({\bf k}) \); the values of ${\bf k}_{c}$
that satisfy this relation for the largest value
of $r(h)$ are at the corners of a hexagonal
Brillouin zone shown in Fig. 3(a). Thus, as
$r(h)$ is decreased the first out-of-plane
corrugations of the colloidal array will occur
at these wavevectors.  When further neighbors
are considered, ${\bf k}_{c}$ can shift to the
edges of the Brillouin zone, as in Figs. 3(b).
The location of these first instabilities
depends on the relative sizes of $K_{1}^{(n)}$. In
the next nearest neighbor truncation, the first
unstable modes occur at the zone edges when

\begin{equation}
\vert K_{1}^{(2)} \vert  > {1 \over 8}
\vert K_{1} \vert
\end{equation}

\noindent as shown in Fig. 3(c).
With repulsive and reasonably short ranged
potentials (such that $K_{1}^{(n>2)} \approx 0$),
the corner and edge instabilities are always the
ones to occur first.  We will restrict ourselves
to these two types of instabilities. The real
space structures of these corrugations are
depicted in Fig. 4.

Since three pairs of corner instability points
are connected via reciprocal lattice vectors,
the height function $f$ is the real part of a
one component complex function. The edge modes
can be represented by  a three component
function. The sums over critical wavevectors
${\bf k}_{c}$ are thus restricted to the stars
of $\pm {\bf k}_{o}$ and ${\bf q}_{i},
(i=1,2,3)$ belonging to the symmetries of the
corner and edge instabilities respectively,

\begin{equation}
{\bf k}_{o} \equiv \pm{4\pi \over 3a}{\bf \hat {y}}
\label{bc}
\end{equation}

\begin{eqnarray}
{\bf q}_{1} &\equiv & {2\pi \over {\sqrt 3} a}
{\bf \hat {x}} \nonumber \\
{\bf q}_{2} &\equiv &{- \pi \over {\sqrt 3} a}
{\bf \hat {x}} - {\pi \over a}{\bf \hat {y}} \nonumber \\
{\bf q}_{3} &\equiv &{- \pi \over {\sqrt 3} a}
{\bf \hat {x}}+{\pi \over a}{\bf \hat {y}} \label{be}
\end{eqnarray}

These two sets of unstable modes transform as
particular irreducible representations of the
space group of the lattice. Since the height
function can be written as linear combinations of
the basis functions of one of these two
irreducible representations (excluding the unit
representation), a continuous transition into
these ordered structures is possible if no cubic
invariant in the free energy
exists\cite{SPT,LDL}. This Landau rule is
not exact in two dimensions
because fluctuations can lead to continuous
transitions even when the cubic term is
present\cite{AA,Zia}. Landau theory nevertheless plays
an essential role in determining the {\it universality class}
of the transition. The functions $f(\pm {\bf k}_{o})$
and $f({\bf q}_{i})$ are identified with $d$ component order
parameters where $d=2,3$ are the dimensionalities
of the irreducible representations\cite{LDL,ED1}.

When unstable modes with wavevectors at zone
boundaries occur, the energy will depend on the
phase of $f$. For example, with ${\bf k}$ and
$-{\bf k}$ pointing to the midpoint of an edge of
the Brillouin zone boundary, careful enumeration
of the sums over wavevectors of the reduced
group, or star, leads to a quadratic term \(
\propto {1\over 2}(f_{k}f_{k} + c.c.) \).

The free energy without phase terms is correct
for wavevectors such that \({\bf k}+{\bf k'}
\neq {\bf G}\) and is sometimes termed the
``incommensurate'' free energy. However, when
the relevant wavevector is is a rational
fraction of the reciprocal lattice vector, the
phase variable enters and a ``commensurate''
free energy obtains\cite{LT}.  Since the
functions $\sum_{n}D_{n}({\bf k})$ for $n \leq
3$ appear to have stable extrema only at corners
or midpoints of edges, the commensurate
wavevectors (\ref{bc}) and (\ref{be}) are the most
important. Thus we expect the system to order
directly into a corrugation with commensurate
wavevector.

We now derive these commensurate energies.
Since the energy was a sum over discrete lattice
sites, the expression (\ref{M0}) contains sums
over $\bf G$.  Thus, at the boundaries of the
Brillouin zone, certain wavevectors can
participate in the ``Umklapp processes'' shown
in Fig. 5.  It is these terms that generate
$\phi$ dependent terms in the energy that lock
the phase. The phase dependence can be
incorporated by generalizing the order
parameter, and allowing for a slow spatial
undulations in both the phase and amplitude,

\begin{equation}
f({\bf k}) = f_{\bf k}(x)e^{i\phi_{\bf k}(x)}, \label{ft}
\end{equation}

\noindent where $f_{\bf k}(x)$ and $\phi_{\bf
k}(x)$ are real positive functions. The
procedure we follow is first fixing the
amplitude and phase of the order parameters,
expressing the energy in terms of these
variables, then allowing slow spatial
modulations within the sample.

At ${\bf k}_{c} = \pm {\bf k}_{o}$ (the corner
instabilities shown in Fig. 3(a)), the
nearest neighbor, ${\bf u}({\bf k})$ independent
part of the commensurate free energy becomes,

\begin{eqnarray}
\bar{\Omega} = {1 \over A}
\left( r(h)+9K_{1}\right)
\vert f({\bf k}_{o}) \vert ^2
+ {1 \over A}\left( 6u +
{81K_{2} \over 4a^2} \right)
\vert f({\bf k}_{o}) \vert ^4 +
+{1 \over A} \left(20v+{243K_{3}
\over 8a^{4}}\right) \vert
f({\bf k}_{o}) \vert ^6 \nonumber \\
+{1 \over A}\left(2v -{81K_{3} \over 8a^4}\right)
cos(6\phi)\vert f({\bf k}_{o})
\vert ^6 + \ldots \label{F1}
\end{eqnarray}

Since \(2v-81K_{3}/8a^4 > 0 \), the six phases
minimizing (\ref{F1}) for the $\sqrt 3 \times \sqrt
3$ state are $\{ \phi_{m}\} = \{(2m+1)\pi /6, \,
m=0,...,5 \}$. They represent three equivalent states
generated by translations in three directions. This
degeneracy in each of these states is doubled by the
inversion $f({\bf k}_{o}) \rightarrow -f({\bf
k}_{o})$.

For edge wavevectors the quadratic terms fix the
phase $\phi$ of the displacement fields at $\phi
= m\pi$, provided higher order terms are
sufficient to stabilize the energy. For these
critical modes, ${\bf k}_{c} = {\bf q}_{i}$ ($i
= 1,2,3$), we have

\begin{eqnarray}
\bar{\Omega} = {1 \over A}
\left( 4K_{1} + 4K^{(2)}_{1} +{r(h) \over 2} \right )
\sum_{i=1}^{3} (1+cos \, 2 \phi_{i} \,)
\vert f_{i} \vert ^2
+ \left({u \over A}+ {4K_{2} \over a^2A} +
{4K_{2}^{(2)} \over 3a^2A} \right)
\sum_{i=1}^{3} \vert f_{i}\vert ^4
cos 4\phi_{i} + \nonumber \\
\left({u \over A}+ {2K_{2} \over a^2A}
+{2K_{2}^{(2)} \over 3a^2A} \right)
 \sum_{i=1}^{3} \left[ 3 \vert f_{i} \vert^2
\vert f_{i+1} \vert^2  cos 2 (\phi_{i}
+\phi_{i+1}) + 3\vert f_{i} \vert^2
(\vert f_{i-1} \vert^2
+ \vert f_{i+1} \vert ^2)
cos 2\phi_{i} \right ] + \ldots \label{F2}
\end{eqnarray}

\noindent where we have set $f_{i} \equiv f({\bf
q}_{i})$ and $\phi_{i} \equiv \phi({\bf
q}_{i})$.  In this case the quadratic term
selects the phase. For \( 4K_{1} + 4K_{1}^{(2)}
+r/2 < 0, \phi = m\pi \) and the
quadratic coefficient in (\ref{F2}) at this
phase is \( 8K_{1} + 8K_{1}^{(2)} +r(h) \).
Comparing (\ref{F1}) with (\ref{F2}) leads to
the condition (\ref{be}) for edge instabilities.

When edge instabilities and phases lock as
described above, the order parameter still
contains three independent real components, $\{
\vert f_{i} \vert \}$ corresponding to three
equivalent directions for the corrugation. In
principle, the quadratic couplings in Eq.
(\ref{F2}) could select a ``triple-{\bf q}''
structure (see Fig. 4(c)), in which all three
components order simultaneously. However, in the
symmetric wall potential case, the ordered state
has a configuration with all the weight in one of
these components which is spontaneously chosen by
the system.  This ``single-${\bf q}$'' state
minimizes the free energy for symmetric wall
potentials and is depicted in Fig. 4(b). This
resembles an observed maze-like
pattern\cite{TO,TOg} and can be an intermediate
structure in the transition to the 2$\Box$ phase.
However, we do not expect antiferromagnetic Ising
behavior (see below) as suggested by
Ogawa\cite{TOg}. For the free energy (\ref{F2}),
a 3-${\bf q}$ state, where all the $\vert
f_{i}\vert $ are equal, is favored only in the
presence of asymmetry in $V(f)$

\section{correspondence of free energy to
continuum statistical models}
We now explicitly allow the order parameter to
acquire a slowly spatially varying modulation,
as defined in (\ref{ft}). This is equivalent to
expanding ${\bf k}$-dependent quadratic
coefficients, $D_{1}({\bf k})$ about ${\bf k}
\simeq {\bf k}_{o}$. With the notation $f_{o}(x)
\equiv f_{{\bf k}_{o}}(x)$ and \( \phi_{o}(x)
\equiv \phi_{{\bf k}_{o}}(x) \), the nearest
neighbor free energy in position space becomes,

\begin{eqnarray}
\bar{\Omega} = {3a^2 \vert
K_{1}\vert\over  4A} \vert
\nabla f_{o}(x) \vert ^2 + {1
\over A}\left( 9K_{1} + r(h) \right ) f_{o}^2(x)
+ {1 \over A} \left( 6u+{81 \over
4a^2}K_{2}\right) f_{o}^4(x) + \nonumber \\
{1\over A} \left( 20v+{243 \over 8a^4}K_{3}
\right) f_{o}^6(x) + {1 \over A}\left( 2v-{81
\over 8a^4}K_{3} \right)cos \,6\phi_{o} (x)
\,f_{o}^6(x) + \ldots \label{F3}
\end{eqnarray}

\noindent At a fixed amplitude which
minimizes $f_{o}$, the expression
describing the energetics of the phase
variable becomes

\begin{equation}
\bar{\Omega} = {3a^2
\vert K_{1} \vert \over  4A}
\vert \nabla \phi_{o}(x)
\vert ^2  f_{o}^2 +{1 \over A}
\left( 2v-{81 \over
8a^4}K_{3} \right) f_{o}^6 \,
cos \,6\phi_{o}(x)
+ \ldots \label{F4}
\end{equation}

This expression has the form of the 2D X-Y model
with six-fold anisotropy as discussed by
Jos\'{e} and others\cite{DRN,JJ}. This model is
expected to order in two steps with an
intermediate region of continuously varying
exponents as the parameters are varied.

When asymmetry exists,

\begin{equation}
\bar{\Omega} = {3a^2
\vert K_{1}\vert \over  4A}
\vert \nabla \phi_{o}(x) \vert ^2
\eta_{o}^2 +{2 \over A}{\bar{\lambda}}
\eta_{o}^3 \, cos\,3
\phi_{o}(x) + \ldots
\end{equation}

\noindent Here, the largest phase locking term
is proportional to $\eta_{o}^3cos \,3\phi_{o}$
To minimize the energy, the phases prefer the
values $\phi_{o} = \{\pm {\pi \over 3}, \pi \}$
when $\bar{\lambda} > 0$ and $\{0, \pm {2\pi
\over 3} \}$ when $\bar{\lambda}< 0$. This
hamiltonian is identical to that of the
continuum 3-state Potts model\cite{SA}. The
behavior of this hamiltonian has been studied
extensively\cite{SA,Zia,MEF,GG}; a {\it
continuous} transition is possible, and the
critical exponents are known\cite{Zia}.

When the edge modes buckle, there exist two
cases.  If a 1-${\bf q}$ state is favored, any
one of three directions is selected, each with
two preferred phases $\phi$. These conditions
are described by a Heisenberg model with quartic
couplings proportional to $ \left(
\sum_{i=1}^{3} \vert f_{i} \vert ^{2}
\right)^{2}$ and $\sum_{i=1}^{3} \vert f_{i}
\vert ^{4}$. The exact critical behavior of this
model in two dimensions is, to the best of our
knowledge, as yet unknown.

In fact, for the particular hamiltonian at hand,
the 3-${\bf q}$ state is energetically preferred
over the 1-${\bf q}$ only when appreciable
symmetry breaking interactions (gravity or
asymmetric plates) are present. The term,

\begin{equation}
\bar{\lambda} \sum_{i=1}^{3} \eta_{1}\eta_{2}\eta_{3}\,
cos(\phi_{i} + \phi_{i+1} - \phi_{i+2})
\end{equation}

\noindent favors the 3-${\bf q}$ or $2 \times 2$
state shown in Fig. 4(c); energy configurations
with any $\eta_{i} = 0$ will be unaffected.  In
the presence of such a term, the degeneracy in
phases of the 3-${\bf q}$ structure is halved.
Now the sets \( \{ \phi_{i} \} = \{(\pi, \pi,
\pi), (0,0, \pi), (0, \pi, 0), (\pi, 0, 0)\}\)
and \( \{(0,0,0), (\pi, \pi, 0), (\pi, 0, \pi ),
(0, \pi , \pi) \} \) are selected when
$\bar{\lambda} > 0$ and $\bar{\lambda} <0$
respectively. A 4-state Potts model (which
also has a continuous transition in $d=2$) is recovered
in this case\cite{ED1,Zia,ANB}. Of course, as
$r(h)$ is decreased further, discontinuous
transitions from the 3-${\bf q}$ to 1-${\bf q}$
corrugations are possible.

The minimum energy configurations in order
parameter space are shown in Fig. 6. In general,
a state described by an equally weighted
superposition of directions, the 3-${\bf q}$
state, has eight possible combinations of the
three phases associated with each order parameter
component that minimize the energy, as shown in
Fig. 6(c). The $\sqrt 3 \times \sqrt 3$ corner
instability is represented by Fig.  6(a) and Fig.
7(a) for the symmetric and asymmetric cases
respectively. For the free energies derived from
our microscopic model, the situation depicted in
Fig. 6(c) does not occur since in the absence of
a symmetry breaking term, the 1-${\bf q}$ state
(Fig. 6(b)) is always preferred.

\section{Experimental Consequences}
In this section, we briefly examine the
experimental signatures expected for each of the
buckled structures above. To begin, we must have
a rough description of the interaction
parameters. The simplest DVLO type potentials
for the sphere-sphere and sphere-wall
interactions in the charge stabilized limit are
the Yukawa and exponential forms respectively,

\begin{equation}
U(r) = {U_{o} \over r}
e^{-\kappa r} \label{I1}
\end{equation}

\begin{equation}
V(f,h)= V_{o}e^{-\kappa(h/2-f)}+
V_{o}e^{-\kappa(h/2+f)} \label{I2}
\end{equation}

In the experiments of Murray et.al.\cite{CAM2},
a layer of $0.3 \mu m$ poly(styrene) sulfonate
spheres were confined in a gap $h \sim 3\mu m$.
To stabilize the structures, low ionic strength
solvents with $\kappa a \sim 0.05$, where
$\kappa$ is the Debye screening length, were
used.  With these estimates we find $\vert K_{i}
\vert \sim 1 \times 10^{-3} erg/cm^2, \, r(h) \sim
0.1 erg/cm^{2}$, and $u(h) \sim 1 erg/cm^{4} $.
Interactions of the form (\ref{I2}) make $r$ and
$u$ very sensitive to the gap spacing $h$ and
very small wedge angles are probably needed to
study these transitions in detail. Although our
analysis has neglected the equilibrium thermal
fluctuations of the spheres in solution, it will
correctly identify the symmetry or ``universality
class'' of the various transitions. Thermal
fluctuations will change the detailed predictions
of mean field theory whenever the mean square
fluctuations of the order parameter amplitude
exceeds its equilibrium value.

Asymmetry effects can be described by the
additional term (\ref{V1}).  If the external
field of concern is gravity \( \lambda =
v_{o}(\rho_{c} - \rho _{o})g \), where $v_{o}$
is the volume of each particle and $\rho_{c}$
and $\rho _{o}$ are the densities of the colloid
and suspending fluid respectively. For
poly(styrene) spheres, $\lambda \simeq 7 \times
10^{-13}\, erg/cm$.

An estimate of the relative displacement of the
entire layer from the center of the gap using
the values in (\ref{shift1}) gives \( \vert
\bar{f}/h \vert \sim 1\times 10^{-8} \) which is
truly negligible. Another measure of the degree
of asymmetry is the change in the critical value
$r^{*}(h)$ where the magnitude $f_{\bf k}$ first
becomes non-zero. At a corner instability
($\sqrt 3\times \sqrt 3$), for \( \lambda <<
K_{1}^{2}/u \),

\begin{equation}
r_{\lambda}^{*}-r_{o}^{*} \equiv
\delta r^{*} \simeq {4 \lambda^{2}
\over 81K_{1}^{2}} \left[ {4a^{2}u\over
24ua^{2}+81K_{2}} -3 \right] \label{shift2}
\end{equation}

\noindent Since $K_{2} > 0$ and $\bar{f}$
depends on $r(h)$, the critical value $r^{*}(h)$
is {\it lowered} by the presence of $\lambda$,
{\it i.e.,} a larger $h^{*}$ is required to buckle the
lattice.

Given that the conditions for equilibrium
ordering are met, there are essentially two
methods currently used to probe these
structures, direct video imaging and
diffraction. When thermal fluctuations are
strong, laser light diffraction patterns can be
analyzed to elucidate average structural
features.

For monochromatic light incident on the layers
(the plates are made transparent), diffraction
peaks are expected at wavevector transfers equal
to the critical wavevectors (\ref{bc},\ref{be}).
The $\sqrt 3 \times \sqrt 3$ structure is
defined by a primitive cell with three times the
area of the original, undistorted lattice. A
three point basis is also superimposed. The
static, zero temperature structure factor for
point, single scatterers in this case is,

\begin{equation}
\delta({\bf k_{\perp}, G}) \delta({\bf -k_{\perp},
G'}) \left[ 2+2\,cos ({\bf k}\cdot{\bf
e}_{1}+k_{z}z_{o}) +2\,cos({\bf k}\cdot({\bf
e}_{1}+{\bf e}_{2})+k_{z}z_{o}) \right]
\end{equation}

\noindent where the 2D reciprocal lattice
corresponding to the larger hexagonal lattice
are,

\begin{equation}
{\{\bf G} \} = \left( {n4\pi \over 3a}{\bf \hat{y}},
\, {m4\pi \over 3a}({\sqrt{3} \over 2}{\bf \hat{x}}+
{1 \over 2}{\bf \hat{y}}) \right)
\end{equation}

\noindent and $z_{o}$ is the spacing between the
two flat lattices, $z_{o} = {3 \over 2}f_{\bf
k_{c}}$.  In a $\sqrt 3 \times \sqrt 3$
structure, in addition to the six spots expected
from the original triangular lattice, we would
see six additional first order peaks at smaller
wavevector transfers and rotated by $30^{o}$
provided $z_{o}$ is large enough. This pattern
is in marked contrast to the asymmetric, three
spot pattern seen in diffraction from a
$2\triangle$ superlattice. We note that {\it
all} diffraction spots in two dimensions are
algebraic singularities \cite{DRN}, rather than
true $\delta$-function Bragg peaks.

The $2 \times 2$ structure has even a larger
primitive cell and a four point basis. The six
diffraction pattern peaks would be even closer
to the origin that in the $\sqrt 3 \times \sqrt
3$ structure and would not be rotated with
respect to the six original peaks.  A pattern
consisting of four spots rectangularly
positioned would be an indication of a $2 \times
2$ superlattice. A similar square pattern
develops when the ordering is $2 \Box$. It would
be interesting see how these patterns evolve as
$r(h)$ is decreased.

In addition to diffraction patterns, video
imaging of the colloidal crystal may reveal the
out-of-plane buckling and the concomitant
in-plane lattice distortions. The in-plane
contractions are due to the \( {\bf e_{i}}
\cdot\Delta_{i} {\bf u} \vert \Delta_{i} f \vert
^2 \) terms in (\ref{E3}) which we have
neglected. To predict any effects in the lattice
spacing as buckling occurs, we minimize the
energy density, $\bar{\Omega}$ with respect to
the $\Delta_{i} {\bf u}$ associated with the
specific corrugations in Fig. 4. The
$\Delta_{i} {\bf u}$ used are the ${\bf k} = 0$
modes expected on symmetry grounds for each of
the structures. A $\sqrt 3 \times \sqrt 3$
buckling is expected to induce a symmetric
contraction of magnitude,

\begin{equation}
{\Delta a \over a} \simeq {3 \over 4a^{2}}
\langle f^{2}({\bf k}_{o}) \rangle
\end{equation}

\noindent The contractions depend on $\langle
f_{\bf k}^{2} \rangle$ which, in turn behaves
according to the relevant model near the
transition with a $\vert r-r^{*} \vert
^{1-\alpha}$ energy-like singularity. Similar
expressions hold for the other buckled
structures. Direct observation of lattice
contractions may reveal the nature of the
buckling transition even if the predicted extra
spots are blurred by thermal fluctuations. It is
not clear how strongly the in-plane
displacements are coupled to the buckling at
larger amplitudes. For instance, individual
fluctuations in $f({\bf x})$ of the charge
stabilized particles can induce a fluctuating
dipole in the ${\bf \hat{z}}$ direction which
would contribute a net repulsion between the
particles tending to expand the lattice.

\section{Summary}
Stability analysis has shown that the buckling
of a layer of confined repelling particles
results from the competition between a bulk
compressional force derived from a chemical
potential and a confining wall force derived
from a repulsive particle-wall potential.  We
have found that the unstable modes have
wavevectors ${\bf k_{c}}$ at the midpoint of
edges and at the corners of the Brillouin zone
boundaries. These instabilities correspond to
the $2 \times 1$ or $2 \times 2$ and the $\sqrt
3 \times \sqrt 3$ structures respectively. In
this analysis, the edge instabilities arose only
when further neighbor interactions were
considered.  The additional restriction
(\ref{w}) was imposed for the in-plane phonons
to remain stable.

The symmetry allowed buckled states have basis
functions belonging to irreducible
representations of the point group. Furthermore,
an expansion of the order parameter coefficients
in the wavevector index ${\bf k}$ yields minima
at ${\bf k_{c}}$. Despite the Landau rule
prohibiting a continuous transition in the
presence of gravity or plate
asymmetry\cite{SPT,LDL,LT}, we find that in this
two dimensional problem, second order transitions
may occur \cite{ED1,GG}.

Provided we allow the order parameter {\it
phase} to be slowly space dependent, the free
energy can be written in a continuum form. A
striking feature is that these hamiltonians
allow for continuous transitions into the
mentioned buckled structures, and are predicted
to have critical behavior corresponding to a X-Y
model with six-fold anisotropy for corner
instabilities, and a Heisenberg model for an
edge instability. Colloids between symmetrical
plates may be the first experimentally realized
nonmagnetic system which can be described by an
X-Y model with six-fold anisotropy. In general,
the edge instability has critical behavior
resembling that of a Heisenberg model with cubic
interactions, although in this particular
problem transitions to 3-${\bf q}$ states are
expected to occur only in the presence of
asymmetry.

Asymmetry from the effects of gravity were found
to be negligible. When asymmetry is expected to
be important continuous transitions are not
precluded by cubic invariants in the free
energy.  In these cases the hamiltonians are in
the 3-state and 4-state Potts universality
classes for the corner and edge instabilities
respectively. It would be interesting to use
differently treated upper and lower plates to
enhance this asymmetry.

The structures enumerated above should be easily
distinguished using light diffraction.  We have
estimated the structure factors and predicted
the diffraction patterns that would result from
these superlattices. It is also possible to
obtain an estimate of the contraction in lattice
spacing due to the buckling. This effect is
described by the coupling term $( {\bf
e}_{i}\Delta_{i} {\bf u}) \vert \Delta_{i}f
\vert ^2$, in Eq. (\ref{E3}).

The main prediction of our theory lies in the
analysis of the types of instabilities that may
occur. We emphasize in conclusion that the
precise mode of buckling and hence which type of
phase transition occurs depend sensitively on
the microscopic parameters in the theory.

\acknowledgements
We are grateful to C. A. Murray, who brought
this problem to our attention, and acknowledge
helpful discussions as well with R. Seshadri, E.
Domany, and J. Cardy. Support from the National
Science Foundation through the Harvard Materials
Research Laboratory grant DMR-91-15491, and a
graduate fellowship for T.C.  is gratefully
acknowledged.

\appendix
\section{One-Dimensional System}
In this section, we consider a related problem:
a one dimensional string of confined particles
as shown in Fig. 8 For a line of regularly
spaced nearest neighbor repelling particles
confined to a cylindrical tube, the free energy
expression equivalent to (\ref{E3}) is,

\begin{eqnarray}
\bar{\Omega} ={(K_{1}+K_{2}) \over 2L}
\sum_{i}^{N}(u_{i+1}-u_{i})^{2}+
{K_{1} \over 2L} \sum_{i}^{N}
\vert \vec{f}_{i+1}-\vec{f}_{i} \vert
^{2} + {K_{2} \over 8a^{2}L}\sum_{i}^{N}
\left( \vert\vec{f}_{i+1} -\vec{f}_{i}
\vert^{2} \right) ^{2}
+ \ldots \nonumber \\
+{1 \over L} \sum_{i}^{N}
\left( {r \over 2} \vert
\vec{f}_{i} \vert ^2 + u (\vert \vec{f}_{i}
\vert ^{2})^{2} +
\ldots \right) \label{E31d}
\end{eqnarray}
\noindent where the length of $N$ particles
is exactly,

\begin{equation}
L = Na \left[ 1+ \sum_{i}^{N}
(u_{i+1}-u_{i}) \right] \label{length}
\end{equation}

\noindent Equation (\ref{E31d}) can be
supplemented with similar terms to include further
neighbor interactions. The parameters $r(d)$
and $u(d)$ may be tuned by changing the pore
diameter, or alternatively, by using a cone
geometry.  In contrast to the two dimensional
sheet, the in-line displacements $u_{i}$ are
scalars and the out-of-line displacements
$f_{i}$ are two component vectors. The harmonic
free energy expression analogous to (\ref{M0})
is,

\begin{equation}
M_{n}(k) = (K_{2}^{(n)}+K_{1}^{(n)})(2-2 cos \, nka)
\end{equation}

\noindent and,

\begin{equation}
\sum_{n}D_{n}^{ij}(k) = \delta_{ij} \left[ r(d) +
\sum_{n}K_{1}^{(n)}(2-2\, cos \,nka) \right]
\end{equation}

\noindent where $i,j$ label two
orthogonal directions perpendicular
to the tube axis.

The flat state in the nearest neighbor case
(\ref{E31d}) is unstable at wavevectors $
k_{c}=\pm \pi / a$. The free energy of a
structure with an instablility at $k_{c}= \pi/a$
becomes,

\begin{equation}
\bar{\Omega} = \left({2K_{1}\over L}+
{r(d) \over 2L}\right) \vert
\vec{f}(k_{c})\vert ^{2} +
\sum_{i=x,y} \left( {r(d) \over 2L}\,
cos \, 2\phi_{i}
- {K_{1}a^{2} \over 2L}
\vert \nabla \phi_{i} \vert ^{2} \right)
f_{i}^{2}(k_{c}) + \ldots
\end{equation}

In one dimension, however, the inclusion of the
next nearest neighbor interactions pushes the
first unstable modes inside the Brillouin zone,
to smaller $k_{c}$. The free energy of these
longer wavelength modes will have higher order
phase locking terms only if $k_{c}a / \pi$ is
rational.  Otherwise, the free energy resembles
that of an isotropic X-Y model,

\begin{equation}
\bar{\Omega} = {1 \over 2L}
\sum_{k_{c}, n}D_{n}(k_{c})
\vert \vec{f}(k_{c}) \vert ^{2} +
{1 \over 4L}\sum_{k_{c},n}
\sum_{i=x,y}n^{2}a^{2} f_{i}^{2}(k_{c})
{\partial^{2}D_{n}(k_{c}) \over \partial k^{2}}
 \vert \nabla \phi_{i}(x) \vert ^{2} + \ldots
\end{equation}

Furthermore, each component of $\vec{f}_{k}$
contains a phase variable. In addition to the
absolute phase, the phase difference, $ \vert
\phi_{x}-\phi_{y}\vert $ also determines the
structure. At zero temperature, the phases
minimizing the nearest neighbor approximation
(\ref{E31d}) are $\phi_{i} = m\pi$ and the phase
differences are $0$ or $\pi$ corresponding to a
linearly polarized wave as shown in Fig. 8a.

If, from the inclusion of further neighbors,
$k_{c}$ is incommensurate, or if the first phase
locking term is very high order, the prefered
phase differences $\vert \phi_{x}-\phi_{y}\vert
$ vary continuously.  For example, when \(
4\vert K_{1}^{(2)} \vert > \vert K_{1} \vert \)
the next nearest neighbor truncation, ($n=1,2$),
yields,

\begin{equation}
cos\,k_c a = {-K_1 \over 4K_1^{(2)} }
\end{equation}
\noindent and,

\begin{equation}
\bar{\Omega} = {1 \over L}\left(
r(d)+2K_{1}+4K_{1}^{(2)}+ {K_{1}^2 \over
4K_{1}^{(2)}} \right) \vert \vec{f}(k_{c}) \vert
^{2} + \sum_{i=x,y} a^{2}f_{i}^{2}(k_{c}) \left(
{K_{1}^{2} \over 2K_{1}^{(2)}L}-
{8K_{1}^{(2)} \over L}\right)
\vert \nabla \phi_{i}(x) \vert ^{2} +\ldots
\end{equation}

The mean field structures available range from
linearly polarized undulations ($\vert
\phi_{x}-\phi_{y} \vert = m\pi$) to circularly
polarized helices ($\vert \phi_{x}-\phi_{y}
\vert = (2m+1)/2 \pi$). At low enough
temperatures, these spontaneously chosen motifs
may be observable.

\begin{figure}
\caption{A confined layer of
repelling colloid particles.}
\end{figure}

\begin{figure}
\caption{Vectors indexing coordination shells
for a triangular lattice.\, $\protect\vert {\bf
a}^{(1)} \vert = a;\, \protect\vert{\bf
a}\protect^\protect{(2)}\protect\vert=\protect
\sqrt{3}a,$ $\protect\vert{\bf
a}^{(2)}\protect\vert=\protect\sqrt{3}a,
\protect\ldots \) }
\end{figure}

\begin{figure}
\caption{The function \( \omega^2 ({\bf k}) =
r(h)+D_1({\bf k})+D_2({\bf k})\) slightly above
and exactly at the transition. The first
Brillouin zone is shown in the inset. Zeros
first occur at a corner instability (a) when
\(\vert K_1^{(2)}\vert > {1 \over 8} \vert K_{1}
\vert \) or (b) edge instability otherwise.}
\end{figure}

\begin{figure}
\caption{Minimum energy structures with
out-of-plane displacements indicated. (a)
$\protect\sqrt{3} \protect\times
\protect\sqrt{3}$ corrugation. (b) $2
\protect\times 1$ structure.  (c) $2
\protect\times 2$ structure. Note that not all
particles have equal displacement magnitudes;
however, $\protect\sum_{\bf x}f({\bf x})=0$. }
\end{figure}

\begin{figure}
\caption{``Umklapp'' processes for corner and
edge instabilities. In the absence of asymmetry,
${\bf G} = 2{\bf q}_{i}$ for the edge modes: the
phases are fixed by the first quadratic term.
For the corner instabilities, ${\bf G'} = 6{\bf
k}_{o}$ is the first Umklapp term. When a cubic
term exists, a lower order Umklapp, $3{\bf
k}_{o} = {\bf G''}$ appears.}
\end{figure}

\begin{figure}
\caption{Order parameter space representation of
model Hamiltonians. Energy minima are shown by
filled circles. (a) The X-Y model with six-fold
anisotropy when corner instabilities arise. (b)
Heisenberg model when a 1-${\bf q}$ state is
chosen within the edge instability subspace.
(c) Heisenberg model when ``cubic'' interactions
dominate, (3-${\bf q}$).}
\end{figure}

\begin{figure}
\caption{Order parameter representation with
asymmetry. (a) 3-state Potts. (b) Heisenberg
model with cubic interactions. The light and dark
points refer to opposite signs of the asymmetry
parameter $\lambda$.}
\end{figure}

\begin{figure}
\caption{A confining geometry of particles in a
conical pore}
\end{figure}

\end{document}